\documentclass[12pt]{iopart}
\usepackage[english]{babel}
\begin{document}
\title{Gauge invariance of the strong field approximation}
\author{A. Bechler and M. \'Sl\c{e}czka}
\address{Institute of Physics, University of Szczecin, Wielkopolska 15, 70-451 Szczecin, Poland}
\ead{adamb@univ.szczecin.pl}

\begin{abstract}
It is shown that strong field approximation (SFA) can be formulated in a gauge invariant manner order by order of the expansion with no need for various partitions of the Hamiltonian in different gauges.
\end{abstract}
Gauge-invariance of the transition probabilities describing interaction of atomic and molecular systems with intense electromagnetic fields is a long-standing problem. The point is that gauge-invariance can be apparently broken by strong field approximation (SFA), which is most commonly used for the semi-analytic calculations of the amplitudes \cite{Milosevic}. For example, it has been shown recently \cite{BauerJ} that ionization rates and energy distributions of strong-field photoionization differ substantially in the Keldysh-Faisal-Reiss theory \cite{Keldysh, Reiss, Faisal} when calculated in the length- and velocity gauges. Dependence of the SFA amplitude on the choice of gauge has raised a question which gauge is most appropriate for a description of a particular ionization process, depending on the atom, molecule or even on the initial state. For instance, it has been shown in \cite{BauerD} that in the case of ionization of negative ions with a ground state of odd parity the length-gauge SFA agrees better with a numerical solution of time dependent Schr\"odinger equation (TDSE). On the other hand, velocity-gauge calculations seem to reproduce quite well experimental results on photodetachment of negative fluorine ions \cite{Reiss1}, whereas to get a similar agreement in the length-gauge would require alteration of measured peak intensity by a factor ca. 1.45.

Whereas gauge invariance of the transition amplitudes (up to a phase factor) and probabilities cannot raise controversies as such, this fundamental property may be violated by various types of approximate calculations. It has been shown in \cite{Faisal1} that the velocity- and length-gauge SFA amplitudes in the dipole approximation are equivalent in all orders of the expansion, provided appropriate initial- and final-state partitions of the Hamiltonian are chosen. These partitions in general differ from the "natural" partition into the unperturbed atomic Hamiltonian and the perturbation describing iteraction with the laser pulse,
\begin{equation}\label{eq:1}
    \hat{H}(t)=\hat{H}_{\rm{at}}+\hat{F}(t),
\end{equation}
where
\begin{equation}\label{eq:1a}
 \hat{H}_{\rm{at}}=\hat{H}_0+\hat{V}_{\rm{at}}
\end{equation}
 is the atomic Hamiltonian. In \Eref{eq:1a} $\hat{H}_0 =\hat{\bi{p}}^2/2$ is the free Hamiltonian with $\hat{\bi{p}}$ - the momentum operator,  $\hat{V}_{\rm{at}}$ is the binding (atomic) potential, and $\hat{F}(t)$ describes interaction with electromagnetic field in dipole approximation, with the property $\hat{F}(t)\rightarrow 0$ for $t\rightarrow \pm\infty$. For strong external fields it is rather the atomic potential which constitutes a perturbation, at least during action of a strong pulse, so that also another partition may be considered as "natural"
\begin{equation}\label{eq:2}
    \hat{H}(t)=\hat{H}_{\rm{F}}(t)+\hat{V}_{\rm{at}},
\end{equation}
where
\begin{equation}\label{eq:3}
    \hat{H}_{\rm{F}}(t)=\hat{H}_0+\hat{F}(t).
\end{equation}
In \cite{Faisal1} \Eref{eq:3} has been chosen as the final-state partition, with the initial-state partition different from \eref{eq:1} and given by (in atomic units)
\begin{equation}\label{eq:4}
    \hat{H}(t)=\hat{H}_1(t)-\dot{\bi{A}}(t)\cdot\hat{\bi{r}},
\end{equation}
where $\hat{\bi{r}}$ is the coordinate operator, $\bi{A}(t)$ is the vector potential in dipole approximation and
\begin{equation}\label{eq:5}
    \hat{H}_1(t)=\frac{[\hat{\bi{p}}+\bi{A}(t)]^2}{2}+\hat{V}_{\rm{at}}+\dot{\bi{A}}(t)\cdot\hat{\bi{r}}.
\end{equation}
The momentum operator in \Eref{eq:5} corresponds to the canonical momentum $\hat{\bi{p}}=-\rmi\nabla$. It has to be distinguished from the kinetic (gauge invariant) momentum $\bi{p}_{\rm{kin}}=\bi{p}+\bi{A}(t)$. Both coincide at asymptotic times, since we assume that the vector potential is then zero.
Wave function fulfilling Schr\"odinger equation with the Hamiltonian \eref{eq:5} has the form \cite{Rzazewski}
\begin{equation}\label{eq:6}
    \Phi(\bi{r},t)=\rme^{-\rmi \bi{r}\cdot\bi{A}(t)}\psi(\bi{r},t),
\end{equation}
where $\psi(\bi{r},t)$ is the atomic electron wave function obeying Schr\"odinger equation with~ $\hat{H}_{\rm{at}}$.

A wider class of gauge transformations, of which the velocity- and length-gauge are particular cases, was considered in \cite{Vanne}
\footnote{The "traditional" minimal coupling velocity gauge obtained by replacement $\hat{\bi{p}}\rightarrow \hat{\bi{p}}+\bi{A}(t)$ is in \cite{Vanne} named as radiation gauge. In the "true" velocity-gauge the interaction Hamiltonian is $\hat{\bi{p}}\cdot\bi{A}(t)$ without the $\bi{A}^2(t)$ term. In this paper the velocity-gauge is understood "traditionally".}.
The main conclusion of \cite{Vanne} is that constructing transition amplitudes describing interaction with strong external fields it is necessary to consider both the gauge and appropriate partition of the Hamiltonian. In general, expansions of the amplitude in different gauges can lead to the same results only in the limit of infinite series.

In this letter we propose an alternative form of SFA expansion using a guiding principle that it is an expansion in powers of atomic potential, so that powers of $\hat{V}_{\rm{at}}$ in every term in the series have to be carefully accounted for. The velocity-gauge Hamiltonian in the dipole approximation has the form
\begin{equation}\label{eq:7}
    \hat{H}_{\rm{V}}=\frac{1}{2}[\hat{\bi{p}}+\bi{A}(t)]^2+\hat{V}_{\rm{at}}=\hat{H}_{\rm{at}}+\hat{F}_{\rm{V}}(t),
\end{equation}
where
\begin{equation}\label{eq:8}
    \hat{F}_{\rm{V}}(t)=\hat{\bi{p}}\cdot\bi{A}(t)+\frac{1}{2}\bi{A}^2(t).
\end{equation}
Transition to another gauge $g$ is implemented by a unitary transformation of the form $\exp[\rmi \chi_g(\bi{r},t)]$. Vector potential $\bi{A}$, scalar potential $\varphi$ and the state vectors $|\psi\rangle$ transform according to
\begin{equation}\label{eq:9}
    \bi{A}_g=\bi{A}-\nabla \chi_g,\qquad \varphi_g=\varphi+\partial_t\chi_g,\qquad |\psi_g\rangle=\rme^{\rmi\chi_g}|\psi\rangle,
\end{equation}
and the transformed Hamiltonian reads
\begin{eqnarray}\label{eq:10}
    \nonumber\hat{H}_g &=\rme^{\rmi \chi_g(\bi{r},t)}\hat{H}_{\rm{V}}\rme^{-\rmi \chi_g(\bi{r},t)}-\frac{\partial \chi_g(\bi{r},t)}{\partial t}\\
    &=\frac{1}{2}[\hat{\bi{p}}+\bi{A}(t)-\nabla\chi_g]^2+\hat{V}_{\rm{at}}-\frac{\partial \chi_g(\bi{r},t)}{\partial t}.
\end{eqnarray}
Hamiltonian of interaction with the external field in the gauge $g$ is
\begin{eqnarray}\label{eq:11}
    \fl\nonumber\hat{F}_g(t)&=\rme^{\rmi\chi_g}\hat{H}_{\rm{at}}\rme^{-\rmi\chi_g}-\hat{H}_{\rm{at}}+\rme^{\rmi\chi_g}\hat{F}_{\rm{V}}\rme^{-\rmi\chi_g}-\frac{\partial \chi_g(\bi{r},t)}{\partial t}\\
    \fl &=\hat{F}_{\rm{V}}(t)-\frac{1}{2}(\hat{\bi{p}}\cdot\nabla \chi_g+\nabla\chi_g\cdot\hat{\bi{p}})-\nabla\chi_g\cdot\bi{A}(t)+\frac{1}{2}\left(\nabla\chi_g\right)^2-\frac{\partial \chi_g}{\partial t}.
\end{eqnarray}
For gauge transformations compatible with the dipole approximation $\chi_g(\bi{r},t)$ is at most linear in $\bi{r}$. Whereas the vector potential should vanish at asymptotic times, it is not in general a necessary requirement for $\chi_g$, as can be seen, for instance, in the case of transformation from the "traditional" velocity-gauge (radiation gauge) to the "true" velocity gauge \cite{Vanne}.

Using partitions of the Hamiltonian defined by \eref{eq:1} and \eref{eq:2} one obtains two forms of integral equations for the time evolution operator $\hat{U}(t,t')$
\numparts
    \begin{eqnarray}\label{eq:12}
        \hat{U}(t,t')=\hat{U}_{\rm{at}}(t,t')-\rmi\int_{t'}^t dt_1\hat{U}(t,t_1)\hat{F}(t_1)\hat{U}_{\rm{at}}(t_1,t'),\label{eq:12a}\\
        \hat{U}(t,t')=\hat{U}_{\rm{F}}(t,t')-\rmi\int_{t'}^t dt_1\hat{U}(t,t_1)\hat{V}_{\rm{at}}\hat{U}_{\rm{F}}(t_1,t'),\label{eq:12b}
    \end{eqnarray}
\endnumparts
where $\hat{U}_{\rm{at}}$ is generated by atomic Hamiltonian \eref{eq:1a} and $\hat{U}_{\rm{F}}$ - by the Hamiltonian \eref{eq:3}.
Transition amplitude from an initial bound electron state $|\phi_{\rm{i}}(t')\rangle$ at time $t'$ to the final continuum state $|\phi_{\rm{f}}(t)\rangle$ (orthogonal to $|\phi_{\rm{i}}\rangle$) at time $t$ has the form
\begin{equation}\label{eq:13}
    M=\langle\phi_{\rm{f}}(t)|\hat{U}(t,t')|\phi_{\rm{i}}(t')\rangle
               =-\rmi\int_{t'}^t dt_1\langle\phi_{\rm{f}}(t)|\hat{U}(t,t_1)\hat{F}(t_1)|\phi_{\rm{i}}(t_1)\rangle,
\end{equation}
where $|\phi_{\rm{i}}(t_1)\rangle$ denotes the initial state propagated to the transient time $t_1$ by $\hat{U}_{\rm{at}}(t_1,t')$, according to \eref{eq:12a}. It has to be noted that due to partition \eref{eq:1}, with $\hat{H}_{\rm{at}}$ as the unperturbed Hamiltonian, the initial state is propagated to transient times {\it always} by the evolution operator $\hat{U}_{\rm{at}}$, generated by $\hat{H}_{\rm{at}}=-\nabla^2/2+\hat{V}_{\rm{at}}$. Gauge dependent objects in expression \eref{eq:13} are the complete evolution operator $\hat{U}(t,t_1)$ and the interaction Hamiltonian $\hat{F}(t_1)$. As can be seen from first part of \Eref{eq:13} transition amplitude is gauge independent (up to constant phase factor), since under the gauge transformation
\begin{equation}\label{eq:14}
    \hat{U}(t,t')\rightarrow\rme^{\rmi \chi_g(t)}\hat{U}(t,t')\rme^{-\rmi \chi_g(t')}.
\end{equation}
For an asymptotic time $t\rightarrow\infty$ the function $\chi_g$ either vanishes or tends to a constant limit (independent of $\bi{r}$), and vanishes for $t'\rightarrow -\infty$ \cite{Vanne}. For a short pulse with the vector potential vanishing beyond the time interval $(t_{\rm{i}},\,t_{\rm{f}})$ it is sufficient to put $t'$ and $t$ as equal to $t_{\rm{i}}$ and $t_{\rm{f}}$ respectively, since beyond this time interval the time evolution is generated by the unperturbed atomic Hamiltonian.

As usual, transition amplitude is expanded in powers of the atomic potential by substituting consecutive iterations of \eref{eq:12b} into the second expression in \eref{eq:13}. First term in the SFA-expansion reads
\begin{equation}\label{eq:15}
    M_0=-\rmi\int_{t'}^t dt_1\langle\phi^{\rm{a}}_{\rm{f}}(t)|\hat{U}_{\rm{F}}(t,t_1)\hat{F}(t_1)|\phi_{\rm{i}}(t_1)\rangle,
\end{equation}
where $|\phi^{\rm{a}}_{\rm{f}}(t)\rangle$ denotes SFA-approximated final state. Final state is usually approximated by a plane wave $|\bi{q}\rangle$
\cite{Milosevic} or by a plane wave distorted by asymptotic Coulomb phase \cite{Faisal2}. In the first case, which is more appropriate eg. for description of detachment from negative ions, $\hat{U}_{\rm{F}}(t_1,t)|\phi^{\rm{a}}_{\rm{f}}(t)\rangle$ is the Volkov state. Approximation of the final state by distorted plane wave should be used in the calculations of ionization of neutral atoms, when final electron moves in the long-range potential of residual ion. The initial and approximated final state are not orthogonal to each other. Using equation fulfilled by $\hat{U}_{\rm{F}}$,
\begin{equation}\label{eq:16}
    -i\frac{\partial}{\partial t_1}\hat{U}_{\rm{F}}(t,t_1)=\hat{U}_{\rm{F}}(t,t_1)[\hat{H}_0+\hat{F}(t_1)],
\end{equation}
substituting the product $\hat{U}_{\rm{F}}\hat{F}$ from \eref{eq:16} into \eref{eq:15} and performing integration by parts one finds
\begin{equation}\label{eq:17}
   \fl M_0=\langle\phi^{\rm{a}}_{\rm{f}}(t)|\hat{U}_{\rm{f}}(t,t')|\phi_{\rm{i}}(t')\rangle-\langle\phi^{\rm{a}}_{\rm{f}}(t)|\phi_{\rm{i}}(t)\rangle
   -\rmi\int_{t'}^t dt_1\langle\phi^{\rm{a}}_{\rm{f}}(t)|\hat{U}_{\rm{F}}(t,t_1)\hat{V}_{\rm{at}}|\phi_{\rm{i}}(t_1)\rangle,
\end{equation}
where also the equation $\rmi\partial_t|\phi_{\rm{i}}(t)\rangle=[\hat{H}_0+\hat{V}_{\rm{at}}]|\phi_{\rm{i}}(t)\rangle$ fulfilled by the initial state was used at intermediate stage of the calculation. Calculation of the amplitude $M_0$ in another gauge $g$, but still with the use of partitions \eref{eq:1} and \eref{eq:2} of the Hamiltonian, gives
\begin{equation}\label{eq:18}
    M^g_0=-i\int_{t'}^t dt_1\langle\phi^{\rm{a}}_{\rm{f}}(t)|\hat{U}_{\rm{F}}^g(t,t_1)\hat{F}_g(t_1)|\phi_{\rm{i}}(t_1)\rangle,
\end{equation}
where $\hat{U}_{\rm{F}}^g(t,t_1)=\rme^{\rmi\chi_g(t)}\hat{U}_{\rm{F}}(t,t_1)\rme^{-\rmi\chi_g(t_1)}$. Substituting $\hat{F}_g(t_1)$ from \eref{eq:11} and using expression for $\hat{U}_{\rm{F}}^g$ one obtains from \eref{eq:18} after simple though little tedious calculation
\begin{eqnarray}\label{eq:19}
    \fl\nonumber M_0^g=-\rme^{\rmi\chi_g(t)}\int_{t'}^t dt_1\frac{\partial}{\partial  t_1}\langle\phi^{\rm{a}}_{\rm{f}}(t)|\hat{U}_{\rm{F}}(t,t_1)\rme^{-\rmi\chi_g(t_1)}|\phi_{\rm{i}}(t_1)\rangle\\
   \nonumber -\rmi\rme^{\rmi\chi_g(t)}\int_{t'}^t dt_1\langle\phi^{\rm{a}}_{\rm{f}}(t)|\hat{U}_{\rm{F}}(t,t_1)\rme^{-\rmi\chi_g(t_1)}\hat{V}_{\rm{at}}|\phi_{\rm{i}}(t_1)\rangle\\
    \fl\nonumber =\rme^{\rmi\chi_g(t)}\langle\phi^{\rm{a}}_{\rm{f}}(t)|\hat{U}_{\rm{F}}(t,t')|\phi_{\rm{i}}(t')\rangle
    -\langle\phi^{\rm{a}}_{\rm{f}}(t)|\phi_{\rm{i}}(t)\rangle\\
    -\rmi\rme^{\rmi\chi_g(t)}\int_{t'}^t dt_1\langle\phi^{\rm{a}}_{\rm{f}}(t)|\hat{U}_{\rm{F}}(t,t_1)\rme^{-\rmi\chi_g(t_1)}\hat{V}_{\rm{at}}|\phi_{\rm{i}}(t_1)\rangle.
\end{eqnarray}
First two terms in \eref{eq:17} and \eref{eq:19} are equal provided that, apart of the property $\chi_g(t')=0$, also $\chi_g(t)=0$ for asymptotic times. This assumption narrows the class of gauge transformations considered in \cite{Vanne}, but still leaves a wide class of transformations of the type $\chi_g(\bi{r},t)=\gamma\,\bi{r}\cdot\bi{A}(t)$, where $\gamma$ is an arbitrary real parameter. However, remaining contributions in \eref{eq:17} and \eref{eq:19}, expressed as time integrals, are essentially gauge dependent, since $\chi_g(t)\neq 0$ for transient times. For strictly periodic long wave trains, such as considered eg. in \cite{Reiss}, first two terms are zero distributions in the limit of asymptotic times, and what remains is a gauge dependent last term in \eref{eq:17}. Using the plane wave for $|\phi^{\rm{a}}_{\rm{f}}\rangle$ and writing $\hat{V}_{\rm{at}}$ as $\hat{H}_{\rm{at}}-\hat{H}_0$ one can then reproduce from \eref{eq:17} Reiss expression for the amplitude~\cite{Reiss}.

We shall show that consecutive terms in the SFA-expansion with the Hamiltonian partitions \eref{eq:1} and \eref{eq:2} are gauge independent if all contributions with given power of the atomic potential are groupped in one term in the expansion. Therefore, the term which is gauge dependent in \eref{eq:17}, as linear in $\hat{V}_{\rm{at}}$, will be absorbed by next term of the "conventional" SFA-expansion, given by
\begin{equation}\label{eq:20}
    M_1=(-\rmi)^2\int_{t'}^t dt_1\int_{t_1}^t dt_2\langle\phi^{\rm{a}}_{\rm{f}}(t)|\hat{U}_{\rm{F}}(t,t_2)\hat{V}_{\rm{at}}\hat{U}_{\rm{F}}(t_2,t_1)\hat{F}(t_1)|\phi_{\rm{i}}(t_1)\rangle.
\end{equation}
Changing the order of integration, using again \eref{eq:16} to express the product $\hat{U}_{\rm{F}}\hat{F}$ and performing integration by parts we obtain
\begin{eqnarray}\label{eq:21}
   \fl\nonumber M_1=\rmi\int_{t'}^t dt_2\langle\phi^{\rm{a}}_{\rm{f}}(t)|\hat{U}_{\rm{F}}(t,t_2)\hat{V}_{\rm{at}}|\phi_{\rm{i}}(t_2)\rangle
   -\rmi\int_{t'}^t dt_2\langle\phi^{\rm{a}}_{\rm{f}}(t)|\hat{U}_{\rm{F}}(t,t_2)\hat{V}_{\rm{at}}\hat{U}_{\rm{F}}(t_2,t')|\phi_{\rm{i}}(t')\rangle\\
   +(-\rmi)^2\int_{t'}^t dt_2\int_{t'}^{t_2}dt_1\langle\phi^{\rm{a}}_{\rm{f}}(t)|\hat{U}_{\rm{F}}(t,t_2)\hat{V}_{\rm{at}}\hat{U}_{\rm{F}}(t_2,t_1)\hat{V}_{\rm{at}}
   |\phi_{\rm{i}}(t_1)\rangle.
\end{eqnarray}
Adding $M_0$ and $M_1$ we see that the gauge dependent term in \eref{eq:17} is cancelled by the first term of \eref{eq:21}, and the last term in \eref{eq:21} is already of the second order in the atomic potential and should be included into next term in the SFA-expansion. Thus
\begin{equation}\label{eq:22}
    M_0+M_1=M^{(0)}+M^{(1)}+\Or(V^2_{\rm{at}}),
\end{equation}
where
\numparts
    \begin{eqnarray}\label{eq:231}
M^{(0)}=\langle\phi^{\rm{a}}_{\rm{f}}(t)|\hat{U}_{\rm{F}}(t,t')|\phi_{\rm{i}}(t')\rangle-\langle\phi^{\rm{a}}_{\rm{f}}(t)|\phi_{\rm{i}}(t)\rangle,\\
\label{eq:231a}
M^{(1)}=-\rmi\int_{t'}^t dt_1\langle\phi^{\rm{a}}_{\rm{f}}(t)|\hat{U}_{\rm{F}}(t,t_1)\hat{V}_{\rm{at}}\hat{U}_{\rm{F}}(t_1,t')|\phi_{\rm{i}}(t')\rangle.\label{eq:231b}
    \end{eqnarray}
\endnumparts
The terms $M^{(0)}$ and $M^{(1)}$ are {\it separately} gauge independent and their sum gives expansion of the transition amplitude up to and including contributions $\Or(V_{\rm{at}})$. The first contribution $M^{(0)}$ contains no atomic potential and describes direct ionization. The term $M^{(1)}$ corresponds to the situation when emitted electron is scattered once by the atomic potential. It provides a simple picture of this process: initial state $|\phi_i(t')\rangle$ at an initial time is propagated to the transient time $t_1$ under the influence of the pulse field, then "encounters" the atomic potential, and after that propagates to a final time again under the action of the pulse field only.

Next contribution to the amplitude can be obtained by substituting second iteration of \eref{eq:12b} into \eref{eq:13}, which gives
\begin{equation}\label{eq:23aa}
    \fl M_2=(-\rmi)^3\int_{t'}^tdt_1\int_{t_1}^tdt_2\int_{t_2}^tdt_3
    \langle\phi^{\rm{a}}_{\rm{f}}(t)|\hat{U}_{\rm{F}}(t,t_3)\hat{V}_{\rm{at}}\hat{U}_{\rm{F}}(t_3,t_2)\hat{V}_{\rm{at}}\hat{U}_{\rm{F}}(t_2,t_1)
    \hat{F}(t_1))|\phi_{\rm{i}}(t_1)\rangle.
\end{equation}
Changing order of integration, using again \eref{eq:16} to express the product $\hat{U}_{\rm{F}}\hat{F}$ and performing integration be parts we get for $M_2$
\begin{eqnarray}\label{eq:25}
\fl\nonumber M_2&=\int_{t'}^tdt_3\int_{t'}^{t_3}dt_2\langle\phi^{\rm{a}}_{\rm{f}}(t)|\hat{U}_{\rm{F}}(t,t_3)\hat{V}_{\rm{at}}\hat{U}_{\rm{F}}(t_3,t_2)\hat{V}_{\rm{at}}
|\phi_{\rm{i}}(t_2)\rangle \\
\fl\nonumber
&+(-\rmi)^2
\int_{t'}^tdt_3\int_{t'}^{t_3}dt_2\langle\phi^{\rm{a}}_{\rm{f}}(t)|\hat{U}_{\rm{F}}(t,t_3)\hat{V}_{\rm{at}}\hat{U}_{\rm{F}}(t_3,t_2)\hat{V}_{\rm{at}}
\hat{U}_{\rm{F}}(t_2,t')|\phi_{\rm{i}}(t')\rangle\\
\fl
&+(-\rmi)^3\int_{t'}^tdt_3\int_{t'}^{t_3}dt_2\int_{t'}^{t_2}dt_1
\langle\phi^{\rm{a}}_{\rm{f}}(t)|\hat{U}_{\rm{F}}(t,t_3)\hat{V}_{\rm{at}}\hat{U}_{\rm{F}}(t_3,t_2)\hat{V}_{\rm{at}}
\hat{U}_{\rm{F}}(t_2,t_1)\hat{V}_{\rm{at}}|\phi_{\rm{i}}(t_1)\rangle.
\end{eqnarray}
First term in \eref{eq:25} is equal to the $\Or(V_{\rm{at}}^2)$ contribution in \eref{eq:21} with opposite sign, and the last term in \eref{eq:25} is of the order $V_{\rm{at}}^3$. Therefore, up to and including contributions $\Or(V_{\rm{at}}^2)$
\begin{equation}\label{eq:26}
    M=M^{(0)}+M^{(1)}+M^{(2)},
\end{equation}
where
\begin{equation}\label{eq:27}
  \fl  M^{(2)}=(-\rmi)^2
\int_{t'}^tdt_3\int_{t'}^{t_3}dt_2\langle\phi^{\rm{a}}_{\rm{f}}(t)|\hat{U}_{\rm{F}}(t,t_3)\hat{V}_{\rm{at}}\hat{U}_{\rm{F}}(t_3,t_2)\hat{V}_{\rm{at}}
\hat{U}_{\rm{F}}(t_2,t')|\phi_{\rm{i}}(t')\rangle,
\end{equation}
and $M^{(0)}$ and $M^{(1)}$ are given by \eref{eq:231} and \eref{eq:231b} respectively. It is easy to check that in an arbitrary order with respect to the atomic potential
\begin{eqnarray}\label{eq:28}
    \fl\nonumber M^{(n)}&=(-\rmi)^n\int_{t'}^tdt_n\int_{t'}^{t_n}dt_{n-1}\ldots\int_{t'}^{t_2}dt_1\\
   \fl &\times\langle\phi^{\rm{a}}_{\rm{f}}(t)|\hat{U}_{\rm{F}}(t,t_n)\hat{V}_{\rm{at}}\hat{U}_{\rm{F}}(t_n,t_{n-1})\hat{V}_{\rm{at}}\ldots
  \hat{V}_{\rm{at}}\hat{U}_{\rm{F}}(t_1,t')|\phi_{\rm{i}}(t')\rangle.
\end{eqnarray}
In the expansion of the transition amplitude
\begin{equation}\label{eq:29}
    M=M^{(0)}+\sum_{k=1}^\infty M^{(k)}
\end{equation}
every term is separately independent on the choice of gauge, where gauge invariance has been achieved by appropriate grouping of terms of the same order in the atomic potential.

We have shown that the $S$-matrix expansion including rescattering contributions to all orders can be formulated in a gauge invariant way
provided that all terms of the same order in atomic potential are groupped together. In this approach every term of the expansion is separately gauge independent. A consequence of groupping together all terms with the same powers of atomic potential is that the first term in the SFA-expansion differs from the conventional one, since part of the latter had to be included into the $O(V_{\rm{at}})$ contribution. In every order of expansion gauge dependent parts from previous and next terms cancel. The new form of the expansion proposed here can be used to calculate transition amplitude in intense fields in the SFA-framework up to an arbitrary order in atomic potential with no necessity to choose a particular gauge.

\end{document}